\documentstyle[pre,aps,epsf,preprint]{revtex}
\begin{document}
\title{
Efficient State Preparation via Ion Trap Quantum Computing and 
Quantum Searching Algorithm
}
\author{
Hao-Sheng Zeng and Le-Man Kuang
}
\address{
Department of Physics, Hunan Normal University, Changsha 410081,
             China
}
\maketitle

\begin{abstract}
 We present a scheme to prepare a quantum state in a ion trap with probability 
 approaching to one by means of ion trap quantum computing and Grover's 
 quantum search algorithm acting on trapped ions.
\end{abstract}

Quantum computing [1-5] is a field at the junction of modern
physics and theoretical computer science. Practical experiments
involving a few
  quantum bits have been successfully performed, and much 
progress has
 been achieved in quantum information theory, quantum error 
correction
 and fault tolerant. Although we still far from having desktop 
quantum
 computers in our offices, the quantum computational paradigm could 
soon
  be more than mere theoretical exercise.

The discovery of Shor's quantum algorithm [6,7] for factoring whole numbers 
into
 prime integer factors was a major milestone in the history of quantum
computing. Another significant quantum algorithm is Grover's quantum
search algorithm [8]. Grover's algorithm does not solve NP-complete problems
in polynomial-time, but the wide range of its applications compensates for
this. The key of performing quantum computations is the implementation of a
 quantum computer, because a quantum computer must meet the requirements:
 the isolation from the environment, the controllable and coherent interaction
 between the qubits, and the high-efficiency interrogation of individual qubits.
 Cirac and Zoller have proposed an attractive scheme for realizing a quantum
 computer, which is easily scalable to an arbitrary number of qubits. Their scheme
  is based on a collection of trapped atomic ions, where each qubit is comprised
 of a pair of the ions' internal states. The trapped-ion system features long
  coherent times of the internal qubit states, while the strongly-coupled
  ion motion allows quantum information to be transferred between different
  ions by using a particular quantized mode of the ions' collective motion.
The crucial step is that the "quantum data bus" must be initialized in its ground state. The basics of
  this scheme have been demonstrated experimentally in a fundamental logic
  gate operating between a motional mode of a single trapped ion and two of the
  ion's internal states[9]. Recently, B.E.King et al. have successfully realized
   cooling the collective modes of motion of two trapped ions into their ground
   state[10]. This research pushed the quantum computation by using trapped ions
   to a new level. In this rapid communication, we show that a variety of 
 multi-phonon
 coherent states in a ion trap can be efficiently prepared by combining ion
 trap quantum computing and quantum search algorithm.

We consider a register which consists of $m$ two-level ions forming a string
confined  in a linear rf trap. The electronic states of ions are denoted by
an   integer $k$,
\begin{equation}
|k\rangle _e=|s_m\rangle \otimes |s_{m-1}\rangle \otimes\cdots \otimes|s_1\rangle
\end{equation}
where $m$ is the number of the qubits in the register, $s_i=0, 1$ and
\begin{equation}
k=s_m\times2^{m-1}+s_{m-1}\times2^{m-2}+ \cdots +s_1\times2^0
\end{equation}

The interaction between the center-of-mass motion and the internal
 electronic state of each ion was mediated by a detuned laser pulse.
  For interaction times much greater than the vibrational period of the
   trap, the effective Hamiltonian for the $j$th ion [11] is
\begin{equation}
\hat{H}^{(j)}_i=\hbar\hat{a}^+\hat{a}\chi(\hat{\sigma}^{(j)}_z+\frac{1}{2})
\end{equation}
where $\hat{\sigma}^{(j)}_z$ is the population inversion for $j$th ion,
$\chi=\eta^2\Omega^2/(m\Delta)$ with $\eta$  being the Lamb-Dicke
 parameter, $\Omega$ being the Rabi frequency for the transitions
  between the two internal states of the ions, and  $\Delta$ being
   the detuning    between the exciting laser pulses and electronic
   transition. We assume that these parameters are the same for all ions.
   We choose the durations  $\tau_j$ of   the standing wave pulse to satisfy
   $\chi\tau_j=2^j\pi/2^m$, then the Hamiltonians $\hat{H}^{(j)}_i$  generate
      a unitary transformation
\begin{equation}
\hat{U}=\exp{(-\frac{2\pi i\hat{a}^+\hat{a}\hat{\gamma}}{2^m})}
\end{equation}
where the electronic operator $\hat{\gamma}$  is defined by
\begin{equation}
\hat{\gamma}=\sum^m_{j=1}(\hat{\sigma}^{(j)}_z+\frac{1}{2})2^{j-1}
=\sum^{N-1}_{k=0}k|k\rangle _e\langle k|
\end{equation}
where $N=2^m$ is the size of the register space.

Let $\hat{\Phi}$  is another electronic operator which is canonically
conjugate to $\hat{\gamma}$,  the relation between the eigenstate
 $|\bar{p}\rangle _e$  of $\hat{\Phi}$  and the eigenstate $|k\rangle _e$
 of $\hat{\gamma}$ is connected by the  following  Fourier transformation
\begin{equation}
|\bar{p}\rangle _e=\frac{1}{\sqrt{N}}\sum^{N-1}_{k=0}\exp{(-2\pi ikpN^{-1})}|k\rangle _e
\end{equation}
where $0\leq p\leq N-1$.

We  assume  that all ions are  in the ground state $|0\rangle _e$ and the vibrational state
  is arbitrary.  Then the initial state of the system is
\begin{equation}
|\psi_{in}\rangle =\sum^{\infty}_{n=0}C_n|n\rangle _{vib}\otimes|0\rangle _e
\end{equation}

Now we perform three steps on above equation: Firstly we apply a sequence
 of  $\pi/2$-pulses on the ions to produce a uniform superposition over all
  possible   electronic energy eigenstates, which is also the state
   $|\bar{0}\rangle _e$ in the Fourier   transform basis. Secondly, the unitary
   interaction in Eq.(4) is implemented to couple the vibrational and electronic
   states (note that $\hat{U}$ does not affect the vibrational state, but only
   displaces the eigenstates of the operator $\hat{\Phi}$).  Finally,
   an inverse Fourier transform is run on the electronic register.
   After these three steps,  the system  approaches the following state[12]
\begin{equation}
|\psi_{out}\rangle =\sum^{\infty}_{n=0}\sum^{N-1}_{k=0}C_{k+nN}|k+nN\rangle _{vib}\otimes|k\rangle _e
\end{equation}

If the vibrational state of the ions is a coherent state
\begin{equation}
|\psi\rangle _{vib}=|\alpha\rangle _{vib}=\exp{(-\frac{|\alpha|^2}{2})}
\sum^{\infty}_{n=0}\frac{\alpha^n}{\sqrt{n!}}|n\rangle _{vib},
\end{equation}
 then Eq.(8) becomes
\begin{equation}
|\psi_{out}\rangle =\sum^{N-1}_{k=0}|\alpha,N,k\rangle _{vib}\otimes|k\rangle _e
\end{equation}
where
\begin{equation}
|\alpha,N,k\rangle _{vib}=\exp{(-\frac{1}{2}|\alpha|^2)}\sum^{\infty}_{n=0}\frac{\alpha^{k+nN}}{\sqrt{(k+nN)!}}
|k+nN\rangle _{vib}
\end{equation}

Eq.(10) is an entangled state, the dimension of the entangled space is $N$.
 If we repeatly perform a series of Grover algorithm on Eq.(10), then the
  probability of finding one certain state $|\alpha,N,k_0\rangle _{vib}\otimes|k_0\rangle _e$ will becomes very large,
   and all the others probability become approximately zero. In order to do
    these, we rewrite Eq.(10) as
\begin{equation}
|\psi_{out}\rangle =|\alpha,N,k_0\rangle _{vib}\otimes|k_0\rangle _e+\sum^{N-1}_{k=0, k\neq k_0}|\alpha,N,k\rangle _{vib}\otimes|k\rangle _e
\end{equation}

Then we  perform the Grover's algorithm [13] on Eq.(12) by two steps.
Firstly, we rotate the marked state $|\alpha,N,k_0\rangle _{vib}\otimes|k_0\rangle _e$ by a
phase of $\pi$ to make Eq.(12) becomes the following form
\begin{equation}
|\psi_{out}\rangle =-|\alpha,N,k_0\rangle _{vib}\otimes|k_0\rangle _e+\sum^{N-1}_{k=0, k\neq k_0}|\alpha,N,k\rangle _{vib}\otimes|k\rangle _e
\end{equation}

  Secondly, we continuously apply the inversion operation  about average
  amplitude to Eq.(13) by the following unitary matrix
\begin{equation}
D_{ij}=\left \{
\begin{array}{ll}
\frac{2}{N} & {\rm if\ } i\neq j\\
\frac{2}{N}-1 & {\rm if \ } i=j
\end{array}
\right.
\end{equation}

After performing the above Grover's algorithm one time, we get
\begin{equation}
|\psi_{out}\rangle _1=a_1|\alpha,N,k_0\rangle _{vib}\otimes|k_0\rangle _e+\sum^{N-1}_{k=0, k\neq k_0}b_1|\alpha,N,k\rangle _{vib}\otimes|k\rangle _e
\end{equation}
where $a_1=\sqrt{N}\sin3\theta$, $b_1=\sqrt{(N-1)/N}\cos3\theta$ with $\sin\theta=1/\sqrt{N}$.

If we repeat the above Grover's algorithm $j$ times on Eq. (12), we can
 obtain that
\begin{equation}
|\psi_{out}\rangle _j=a_j|\alpha,N,k_0\rangle _{vib}\otimes|k_0\rangle _e+\sum^{N-1}_{k=0, k\neq k_0}b_j|\alpha,N,k\rangle _{vib}\otimes|k\rangle _e
\end{equation}
where $a_j=\sqrt{N}\sin[(2j+1)\theta]$  and
$b_j=\sqrt{(N-1)/N}\cos[(2j+1)\theta]$.

 From Eq.(16) we can see that if we repeat Grover's algorithm
 $T=(\pi-2\theta)/4\theta $ times, then
 we have $b_T=0$. If we now make a  measurement on
 quantum states of the  register, we can gain the state $|k_0\rangle _e$
 with probability approaching to one. After the measurement,   the vibrational
 state of ions becomes
\begin{equation}
|\alpha,N,k_0\rangle _{vib}=\exp{(-\frac{1}{2}|\alpha|^2)}
\sum^{\infty}_{n=0}\frac{\alpha^{k_0+nN}}{\sqrt{(k_0+nN)!}}|k_0+nN\rangle _{vib}
\end{equation}

The above equation is just the $k_0$-th eigenstate of operator $\hat{a}^N$ . If we just want
 to get a Schr\"{o}dinger-cat state, we need only  one ion ($m=1$, $N=2$).
In fact,  when t $k_0=0$, we   can get the even Schr\"{o}dinger-cat state
 $|\alpha\rangle +|-\alpha\rangle $, when $k_0=1$, we can get the  odd Schr\"{o}dinger-cat state
$|\alpha\rangle -|-\alpha\rangle $. If we want to get four-phonon coherent states,   we need only
 two ions, i.e., $m=2$, and $N=4$.  By selecting  $k_0=0,1,2,3$, we can
  get the  following four-phonon coherent states[14]:
\begin{equation}
|\alpha\rangle _1=|\alpha\rangle +|-\alpha\rangle +|i\alpha\rangle +|-i\alpha\rangle  \hspace{1cm} (k_0=0)
\end{equation}

\begin{equation}
|\alpha\rangle _2=|\alpha\rangle -|-\alpha\rangle -i|i\alpha\rangle +i|-i\alpha\rangle \hspace{1cm}  (k_0=1)
\end{equation}

\begin{equation}
|\alpha\rangle _3=|\alpha\rangle +|-\alpha\rangle -|i\alpha\rangle -|-i\alpha\rangle \hspace{1cm}  (k_0=2)
\end{equation}

\begin{equation}
|\alpha\rangle _4=|\alpha\rangle -|-\alpha\rangle +i|i\alpha\rangle -i|-i\alpha\rangle \hspace{1cm}  (k_0=3)
\end{equation}

It is clear that by means of this method, we can get all the eigenstates
 of operator  $\hat{a}^{2^m}$ (m being the number of ions).

In summary we provide a method which can produces multi-phonon coherent
 states with a completely successful probability. If we do not use Grover's
  algorithm, but measure immediately the electronic state in Eq.(12), then
    we obtain the state  $|k_0\rangle _e$ with only a very less probability
   when $N$  is very  large.    We must note that
   Grover's algorithm can only repeat integer times, but $T=(\pi-2\theta)/4\theta$
     is not always an integer. Therefore we can only repeat Grover's
      algorithm by an integer nearest to $T$. Thus $b_T$ is not completely zero,
       we get the electronic state $|k_0\rangle _e$ with probability only approaching
       to one but not exact one.

This work was supported by  NSF of China, the Excellent Young-Teacher 
Foundation of the Educational Commission of China, ECF, STF and NSF of Hunan Province. .

\end{document}